\documentclass[preprint]{revtex4-1}
\usepackage[USenglish]{babel} 
\usepackage[T1]{fontenc}
\usepackage[ansinew]{inputenc}

\usepackage{lmodern}

\usepackage{graphicx}
\usepackage{amsmath}
\usepackage{amsthm}
\usepackage{amsfonts}

\begin{document}

\title{CP Violation in Hierarchical Yukawa Models}
\author{R.D. Peccei}
\affiliation{Department of Physics and Astronomy\\University of California at Los Angeles\\Los Angeles, California 90095}
\date{\today} 
\begin{abstract}
\noindent Because 4-dimensional CP is a good symmetry of many higher-dimensional theories, this  suggests the possible existence of an universal CP-violating phase originating from the process of compactification. Such a phase, if it existed, would not be easy to uncover since the phases in Yukawa matrices are not simply related to the observed CKM phase $\delta$. Hierarchical Yukawa models, of the type arising in F-theory GUT models provide an interesting exception. Recently, Heckman and Vafa studied a particular F-theory GUT model with hierarchical Yukawa matrices with complex phases of O(1) and showed, by examining the Jarlskog invariant, that this model leads to $\sin\delta\sim$ O(1). A more detailed examination of the model, although confirming their results, is seen to be also compatible with having a phase $\delta_o=\pi /3$ imprinted on the Hermitian Yukawa matrices, leading to $\sin \delta\simeq\sin\delta_o$.
\end{abstract}
\maketitle

\section{Introduction}

It is well known that for CP to be violated the Lagrangian describing the theory must contain complex structures. This can be understood heuristically as follows. Under CP operators get replaced by their Hermitian adjoints: $O(\vec{x},t)\to O^{\dagger}(-\vec{x},t)$. However, because Lagrangians are Hermitian, a Lagrangian containing the operator O has the structure:
\begin{equation}
{\cal L}= aO+ a^*O^\dagger,
\end{equation}
where $a$ is a c-number. It follows thus that a Lagrangian is invariant under CP only if $a=a^*$. So to have CP violated (or, T violated) one must have complex phases in the theory.

\noindent It is reasonable to ask what is the origin of this complexity. Could it be that, in the end, all CP-violating phenomena originate from some simple underlying phenomena? If so, is there perhaps some primordial phase responsible for CP violation in nature? There are grounds to speculate along these lines in higher dimensional theories. One knows that, in general classes of higher dimensional theories, 4-dimensional CP can be embedded as a discrete subgroup of the gauge group associated with these theories. \cite{DLM} \cite{CKN} 

\noindent Let me elaborate briefly on this point. In general, 4-dimensional CP is a good symmetry of any higher-dimensional theory in which fermions and anti-fermions sit in the same representation. An example is provided by 10-dimensional heterotic string theory where fermions and anti-fermions are both in the $E_8$ adjoint representation. Because Charge Conjugation changes $\Psi\to\bar{\Psi}$, C is equivalent to a gauge rotation. Furthermore, in these higher-dimensional theories one can consider ordinary 3-dimensional Parity, which involves the coordinate reversal $\vec{x}\to-\vec{x}$, as being part of a higher dimensional Lorentz transformation. This is easily understood by noticing that one can equivalently think of 3-dimensional Parity as the product of an inversion, times a rotation by $\pi$ in the orthogonal plane:
\begin{equation}
 \vec{x}\to-\vec{x}\equiv \{x_1\to -x_1~;~{\cal R}_{x_2x_3}(\pi)\}.
 \end{equation}
 If one has more than 3 spatial dimensions, then 3-dimensional Parity is part of a higher-dimensional Lorentz transformation. For example,
 \begin{equation}
 \{\vec{x}\to-\vec{x}~:~y\to -y\}\equiv \{{\cal R}_{x_1y}(\pi)~;~{\cal R}_{x_2x_3}(\pi)\}.
 \end{equation}
 
\noindent In higher-dimensional theories where 4-dimensional CP is a good symmetry, CP-violating effects must arise as the result of the compactification from the higher-dimensional space to 4d-space. Thus, in principle, one may be able to compute the resulting 4d CP-violating phases from the underlying geometry. In particular, the complexity which gives rise to the observed CP-violation in the K- and B-system may indeed originate from a simple universal geometric phase. A guess for such an ur-phase is that it could simply be: $\delta_0=\frac{2\pi}{N_{\rm{gen}}}$, which is first non-trivial for $N_{\rm{gen}}=3$.

\noindent Although it is fun to speculate in this way, reality is much more complex. Even if some universal phase existed, its elucidation will not be easy. One of the difficulties is that the number of observables does not match the number of parameters in the Lagrangian. For the discussion that follows, it is sufficient to consider the case of Hermitian Yukawa matrices. This is because, by the polar decomposition theorem, \cite{FJ} any arbitrary Yukawa matrix can be written as a product of a Hermitian matrix and a unitary matrix. Since the unitary matrix can be absorbed through a redefinition of the right-handed quarks, effectively it suffices to study the case of Hermitian Yukawa couplings.

\noindent For three generations, each of the Yukawa matrices $Y^i=Y^{i\dagger}$ ($i=u,d$) is a function of six real parameters and three complex phases. So, altogether, there are 12 real parameters and 6 phases at the Lagrangian level. Experimentally, what is observable are the 6 quark masses and the four parameters in the CKM matrix, \cite{CKM} 3 angles and one phase- ten parameters in total. So it is important to understand  what to look for.

\noindent In this paper we will study the question of the possible existence of an universal phase in the context of a specific model. Although our results are, at first sight, negative, they are useful because they illustrate how difficult it is in practice to arrive at an answer to this question. Indeed, as we shall see, room is left open for an alternative interpretation.

\section{Generalities}

To begin with, it is useful to discuss some well known generalities to set the stage for our considerations. The Hermitian Yukawa mass matrices for the three up and down quarks, $Y^u$ and $Y^d$, are diagonalized by unitary matrices $U^u$ and $U^d$, respectively:
\begin{equation}
U^{u\dagger}Y^uU^u=Y^u_{\rm{Diag}}~~;~~U^{d\dagger}Y^dU^d=Y^d_{\rm{Diag}}
\end{equation}
In Appendix A, following Rasin, \cite{Rasin}, we compute explicitly the matrices $U^i$, with $i=(u,d)$, that diagonalize the Yukawa matrices $Y^i$ in terms of the nine parameters that enter in each of these matrices (six real couplings and three phases). The matrices $U^i$ depend on three real angles and three phases and, in Appendix A, we show that they can be written as
\begin{equation}
U^i=P(0,\delta^i_1,\delta^i_2)V(\theta^i_{23},\theta^i_{13},\theta^i_{12};\delta^i_3)
\end{equation}
Here the matrices P and V (defined in Appendix A) are respectively a phase matrix and a $3 \times 3$ unitary matrix which has the form of the CKM matrix \cite{CKM} written in the "standard" Chau-Keung parametrization. \cite{CK} Because $Y^i_{\rm{Diag}}$ is a function of three eigenvalues, we see that in diagonalizing the Yukawa matrices, the number of parameters is preserved, as it must be.

\noindent Where parameter reduction occurs is in computing the CKM matrix itself. By definition, this matrix is given by the product of $U^{u\dagger}$ and $U^d$ and, obviously, is a $3 \times 3$ unitary matrix. As such, it depends on three real angles and six phases. However, as is well known, five of these phases can be absorbed into redefinitions of the left-handed u- and d-quarks, leaving only one physical phase. That is, one can write:
\begin{equation}
U^{u\dagger}U^d=P(0,\alpha_1,\alpha_2)U_{CKM}(\theta_{23},\theta_{13},\theta_{12};\delta)P(\alpha_3,\alpha_4,\alpha_5)
\end{equation}
Using Eq.(5) we see that
\begin{equation}
U^{u\dagger}U^d=V^{\dagger}(\theta^u_{23},\theta^u_{13},\theta^u_{12};\delta^u_3)P(0,\delta^d_1-\delta^u_1,\delta^d_2-\delta^u_2)V(\theta^d_{23},\theta^d_{13},\theta^d_{12};\delta^d_3).
\end{equation}
Thus $U^{u\dagger}U^d$ is a function of the six angles $\theta^i_{23},\theta^i_{13}$ and $\theta^i_{12}$ and the four phases $\delta_1= \delta^d_1 -\delta^u_1, \delta_2=\delta^d_2-\delta^u_2,\delta^u_3$ and $\delta^d_3$. It is straightforward to extract $U_{CKM}$  from Eqs. (6) and (7), in whatever desired parametrization one decides to choose. However, as we illustrate in Appendix B, the process is rather convoluted. In this Appendix we carry out this process explicitly  for the case where the CKM matrix takes the "standard form" $U_{CKM}=V$. As  can be seen from Appendix B, and as we will discuss in more detail later on in this paper, in general, the measured CKM phase $\delta$ is not only a function of the four CP-violating phases entering in Eq. (7) but also depends on all the other six mixing angles in this equation. Thus it is difficult to gain any insights directly.

\noindent Rather than focusing on $U_{CKM}$ it is useful instead to examine the Jarlskog invariant, \cite{J} which characterizes CP violation in a parametrization independent way. This invariant is defined through the equation
\begin{equation}
J\Sigma_{\gamma k}\epsilon_{\alpha\beta\gamma}\epsilon_{ijk}= Im [U_{\alpha i}U_{\beta j} U^*_{\alpha j}U^*_{\beta i}].
\end{equation}
It is easy to check that J is independent of the phase matrices entering in Eq. (6), so that indeed it does not depend on the parametrization used. Of course, it will be given by different explicit functions of the three angles and one phase chosen to describe $U_{CKM}$. In particular, if we choose the standard parametrization $U_{CKM}=V$ then :
\begin{equation}
J=Im[U_{11}U_{22}U^*_{12}U^*_{21}]=\cos^2\theta_{13}\cos \theta_{12}\cos\theta_{23}\sin\theta_{13}\sin\theta_{12}\sin\theta_{23}\sin\delta.
\end{equation}
Experimentally, \cite{fit} one finds that $J=(3.09\pm0.11)\times 10^{-5}$, which is a small number. However, this number is small not because the phase $\delta$ is small, but because the mixing angles $\theta_{ij}$ are small. Indeed, the best fit of the CKM phase $\delta$ gives \cite{fit}
\begin{equation}
\delta=(69.7\pm2.9)^{o}
\end{equation} 

\noindent Following Wolfenstein \cite{wolf} it has become conventional to expand these angles in powers of the Cabibbo angle, taking
\begin{equation}
\sin\theta_{12}=\lambda~;~ \sin\theta_{23}=A\lambda^2~;~ \sin\theta_{13}e^{-i\delta}=A\lambda^3(\rho-i\eta). 
\end{equation}
Here $\lambda=\sin\theta_C\simeq 0.22$ serves as an expansion parameter and $A,~\rho,$ and $\eta$ are parameters of O(1). Using this approximate parametrization of the mixing angles entering in V, one finds
\begin{equation}
J\simeq A\lambda^6\eta,
\end{equation}
which makes it  clear that J is small, not because $\sin\delta$ is small but because the family mixing- the factor $\lambda^6$ above- is small.

\noindent One can show that J is related to the commutator of the Yukawa matrices. \cite{J} Defining $C=-i[Y^u,Y^d]$, one finds
\begin{equation}
{\rm{Det}}~C=2\Delta J,
\end{equation}
where
\begin{equation}
\Delta=(y^u_3-y^u_2)(y^u_3-y^u_1)(y^u_2-y^u_1)(y^d_3-y^d_2)(y^d_3-y^d_1)(y^d_2-y^d_1).
\end{equation}
Because the eigenvalues of the Yukawa matrices are hierarchical, $y^i_3>>y^i_2>>y^i_1$, the eigenvalue difference function $\Delta$ itself is quite hierarchical:
\begin{equation}
\Delta\simeq [y^u_3y^d_3]^3\left(\frac{y^u_2}{y^u_3}\right)\left(\frac{y^d_2}{y^d_3}\right).
\end{equation}
Hence, Det C is doubly suppressed by hierarchical factors:
\begin{equation}
{\rm {Det}}~C\sim \{\lambda^6\left(\frac{y^u_2}{y^u_3}\right)\left(\frac{y^d_2}{y^d_3}\right)\}[y^u_3y^d_3]^3 \sin\delta.
\end{equation}
Even though the Yukawa matrices for the u- and d-quarks nearly commute, the CP-violating phase $\delta$ meausured experimentally is not suppressed.

\section{Hierarchical Yukawa Models in F-Theory GUTs}
Through the years, many theoretical attempts have been made to construct hierarchical models for the quark mass matrices. A typical example is provided by Froggatt- Nielsen type models \cite{FN} where the Yukawa matrices have the form:
\begin{equation}
Y_{ij}= c_{ij}\epsilon^{a_i+a_j}.
\end{equation}
Here $c_{ij}$ are coefficients of O(1), while $\epsilon<<1$ is a small parameter. The quantities $a_i$ are, so called, Froggatt- Nielsen charges with integer values (e.g. $a_i=\{4,2,0\}$) and they serve to provide a hierarchical structure for the Yukawa matrices. An interesting question in these hierarchical models is the following. If the model reproduces the observed hierarchical pattern in the quark masses and gives a CKM matrix with the right hierarchy, does it follow that $\sin\delta$ will be of O(1) if the coefficients $c_{ij}$ have Arg $c_{ij}$ of O(1)?

\noindent This question has been answered in the affirmative recently by Heckman and Vafa \cite{HV} in the context of a class of F-theory GUT models. These models \cite{f} are higher dimensional theories where 7-branes are wrapped on complex surfaces (S, S',...), which have different gauge groups ($G_S,G_{S'},...$) residing on them. Matter lives on matter curves at the intersection of these surfaces ($\Sigma=S\cap S'$) and an index theorem determines the number of families on $\Sigma$. In these theories the Yukawa couplings arise from the intersection of three matter curves and, to a first approximation the mass matrices are rank one. Including corrections, both $Y^u$ and $Y^d$ are hierarchical. 

\noindent Heckman and Vafa \cite{HV} find the following interesting hierarchical Yukawa patterns in the theory they considered:
\begin{equation}
Y^u \sim \left(\begin{array}{ccc} \epsilon^8_u&\epsilon_u^6&\epsilon_u^4\\\epsilon^6_u&\epsilon_u^4&\epsilon_u^2\\\epsilon^4_u&\epsilon_u^2&1 \end{array}\right)~~;~~
Y^d \sim \left(\begin{array}{ccc} \epsilon^5_d&\epsilon_d^4&\epsilon_d^3\\\epsilon^4_d&\epsilon_d^3&\epsilon_d^2\\\epsilon^3_d&\epsilon_d^2&1 \end{array}\right)
\end{equation}
These matrices lead to a hierarchy of eigenvalues
\begin{equation}
y_1^u:y_2^u:y_3^u \sim \epsilon_u^8:\epsilon_u^4:1~~;~~y_1^d:y_2^d:y_3^d \sim \epsilon_d^5:\epsilon_d^3:1
\end{equation} which reflect the hierarchy of quark masses seen in nature in both the u- and d-sector.
Although these Yukawa matrices do not quite have a Froggatt- Nielsen structure, the matrices $YY^{\dagger}$ do and one finds:
\begin{equation}
[Y^uY^{u \dagger}]_{ij}=\epsilon_u^{a_i +a_j} ~~a_i=(4,2,0)~~;~~[Y^dY^{d \dagger}]_{ij}=\epsilon_d^{b_i +b_j} ~~b_i=(3,2,0).
\end{equation} 
These matrices are diagonalized by matrices $U^u$ and $U^d$ which have the form \cite{HV2}
\begin{equation}
[U^u]_{ij} \sim \epsilon_u^{|a_i-a_j|}~~;~~[U^d]_{ij} \sim \epsilon_d^{|b_i-b_j|}
\end{equation}
Taking $\epsilon_u\simeq\epsilon_d\simeq\lambda$, a simple calculation shows that $U^{u \dagger}U^d$ has the observed hierarchical form:
\begin{equation}
U^{u \dagger}U^d \sim\left(\begin{array}{ccc}1&\lambda&\lambda^3\\ \lambda&1&\lambda^2\\
\lambda^3&\lambda^2&1 \end{array}\right).
\end{equation}

\noindent To examine CP violation in this model, Heckman and Vafa \cite{HV} compute the commutator of the two Yukawa matrices and, again in the approximation where $\epsilon_u\simeq\epsilon_d\simeq\lambda$, find that
\begin{equation}
C\sim \left(\begin{array}{ccc}\lambda^7&\lambda^5&\lambda^3\\ \lambda^5&\lambda^4&\lambda^2\\
\lambda^3&\lambda^2&\lambda^4 \end{array}\right).
\end{equation}
To compute the Jarlskog invariant J one must compute Det C. This is a little tricky to do because of cancellations among different terms of the same order in the determinant. At any rate, Heckman and Vafa \cite{HV} find:
\begin{equation}
{\rm{Det~ C}}= 2\Delta J \sim\lambda^{13}.
\end{equation}
Because in the model
\begin{equation}
\Delta\simeq [y^u_3y^d_3]^3\left(\frac{y^u_2}{y^u_3}\right)\left(\frac{y^d_2}{y^d_3}\right)\sim \lambda^7,
\end{equation}
it follows that 
\begin{equation}
J=\cos^2\theta_{13}\cos \theta_{12}\cos\theta_{23}\sin\theta_{13}\sin\theta_{12}\sin\theta_{23}\sin\delta \sim \lambda^6,
\end{equation}
which predicts for this theory that $\sin\delta \sim O(1)$. 

\noindent We see that, in the case of this F-theory GUT with Yukawa couplings which produce the right hierarchy for masses and mixing angles, the assumption that the phases that enter in the Yukawa mass matrices are of O(1) [Arg $Y_{ij} \sim$ O(1) for $i\neq j$] predicts that the observable CP-violating phase in the CKM matrix is also of O(1). The lesson learned from this example appears to be that the compactification does not produce an ur-CP violating phase. Rather, it produces a hierarchy of couplings each with phases of O(1) in the Yukawa matrices which, in turn, gives rise to a CKM phase of the same order.

\section{A More Detailed Examination}

Although the conclusion arrived by Heckman and Vafa is correct, a more detailed examination of what happens in this model of hierarchical Yukawa couplings is informative. This is the purpose of this Section, with the technical details relegated to Appendix A and Appendix B. As we mentioned in the preceeding Section, the Hermitian Yukawa matrices $Y^i$, which depend on six real couplings and three phases, are diagonalized by unitary matrices $U^i$ which depend on three real angles and three phases, as detailed in Eq. (5). These unitary matrices are found in all generality in Appendix A, but simplify considerably when the real couplings in $Y^i$ are hierarchical. In this case, using the results of Appendix A, one finds that the three mixing angles, in leading order in an expansion in $\epsilon^i$, are given by:
\begin{equation}
\theta^i_{23}= y^i_{23}~~;~~ \theta^i_{13}=x^i_{13}~~;~~ \theta^i_{12}=\frac{x^i_{12}}{x^i_{22}}.
\end{equation}
Here the parameters $x^i_{ab}$ are parameters in the auxiliary "standard form" matrix $Y^i_{\rm{SF}}$ and, in leading order, one finds
\begin{equation}
x^i_{22}=y^i_{22}-(y^{i}_{23})^2
\end{equation}
and
\begin{equation}
\left(\begin{array}{c} x^i_{12}e^{i\alpha^i_{12}} \\x^i_{13}e^{i\alpha^i_{13}} \end{array}\right)=\left(\begin{array}{cc} 1&y^i_{23}\\-y^i_{23}&1 \end{array}\right)\left(\begin{array}{c} y^i_{12}e^{i(\gamma^i_{12}+\gamma^i_{23})}\\y^i_{13}e^{i\gamma^i_{13}} \end{array}\right).
\end{equation}
Here the phases $\alpha^i_{12}$ and $\alpha^i_{13}$  determined from the above equation are related to the phases $\delta^i_1,~\delta^i_2$ and $\delta^i_3$ in $U^i$ and one finds:
\begin{equation}
\delta^i_1=\gamma^i_{23} -\alpha^i_{12}~~;~~\delta^i_2=-\alpha^i_{12}~~;~~\delta^i_3=\alpha^i_{12} -\alpha^i_{13}.
\end{equation}

\noindent Because the hierarchies in $Y^u$ and $Y^d$ shown in Eq. (18) are slightly different, the results for the mixing angles and phases in the u- and d-sector also differ. For the u-sector, since  $y^u_{12}$ and $y^u_{23}y^u_{13}$ are of O($\epsilon^6_u$) the phase $\alpha^u_{12}$ depends in detail on the structure of $Y^u$. It follows from Eq. (29) that $x^u_{13}=y^u_{13}$ and $\alpha^u_{13}=\gamma^u_{13}$, but
\begin{equation}
x^u_{12}e^{i\alpha^u_{12}}=y^u_{12}e^{i(\gamma^u_{12}+\gamma^u_{23})}+ y^u_{23}y^u_{13}e^{i\gamma^u_{13}}
\end{equation}
For the d-sector, on the other hand, $y^d_{12}\sim\epsilon_d^4$ while $y^d_{23}y^d_{13}\sim \epsilon_d^5$ so, in leading order in $\epsilon_d$,  $x^d_{13}=y^d_{13}$ and $x^d_{12}=y^d_{12}$ while the phases $\alpha^d_{13}$ and $ \alpha^d_{12}$ are directly related to the phases appearing in $Y^d$:
\begin{equation}
\alpha^d_{13}=\gamma^d_{13}~~;~~\alpha^d_{12}=\gamma^d_{12} +\gamma^d_{23}.
\end{equation}
Hence
\begin{equation}
\delta^d_1=-\gamma^d_{12}~~;~~\delta^d_2=-\gamma^d_{12}-\gamma^d_{23}~~;~~\delta^d_3=\gamma^d_{12} +\gamma^d_{23} -\gamma^d_{13}.
\end{equation}
Finally, as already anticipated in Eq. (21), one has
\begin{equation}
\theta^u_{12}\sim\epsilon_u^2~~;~~\theta^u_{23}\sim\epsilon_u^2~~;~~\theta^u_{13}\sim\epsilon_u^4~~;~~\theta^d_{12}\sim\epsilon_d~~;~~\theta^d_{23}\sim\epsilon_d^2~~;~~\theta^d_{13}\sim\epsilon_d^3.
\end{equation}
\noindent
\noindent It remains to extract the CKM matrix from Eqs. (6) and (7).  Appendix B details the procedure for doing this in the specific case where the CKM matrix is parametrized in the "standard form" $U_{CKM}=V$, but where the matrices $U^u$ and $U^d$ are general. These general results simplify considerably in the case where there is a hierarchy. In this case, it is straightforward to compute the  parameters in the CKM matrix, including the phase $\delta$, in terms of the mixing angles and phases entering in $U^u$ and $U^d$ detailed above. Before displaying the final results, it is useful to describe qualitatively the steps for computing $U_{CKM}$ and the simplifications that occur at each stage for the hierarchical case under consideration. Using the explicit form for the CKM matrix V, Eq. (7) reads:
\begin{eqnarray}
U^{u\dagger}U^d&=& R_{12}^T(\theta^u_{12})P(0,0,\delta^u_3)R_{13}^T(\theta^u_{13})P(0,0,-\delta^u_3)R_{23}^T(\theta^u_{23})P(0,\delta_1,\delta_2) \nonumber \\~~~~~~&\times &R_{23}(\theta^d_{23})P(0,0,\delta^d_3)R_{13}(\theta^d_{13})P(0,0,-\delta^d_3)R_{12}(\theta^d_{12}),
\end{eqnarray}
where the various rotation matrices $R_{ij}(\theta)$ are detailed in Apppendix A.
To transform this expression into the form of Eq. (6), so as to extract  the CKM matrix $V(\theta_{23},\theta_{12},\theta_{12};\delta)$, one goes through four steps, which are described in detail in Appendix B and summarized below

\noindent {\it{Step i:}}

\noindent The two $R_{23}$  matrices and the central phase matrix are combined to yield a new angle $\phi_{23}$ and two other phases $\gamma_1$ and $\gamma_2$. For the hierarchical case, in leading order in $\epsilon_u$ and $\epsilon_d$, one has
\begin{equation}
\gamma_1=\delta_1~~;~~\phi_{23}e^{i(\gamma_2-\delta_1)}=\theta^d_{23}-\theta^u_{23}e^{i(\delta_2-\delta_1)}.
\end{equation}

\noindent{\it{Step ii:}}

\noindent After some rearrangement of the phase matrices, the product of the two $R_{13}$ matrices and two complex conjugate phase matrices with $R_{23}(\Phi_{23})$ gives a particular parametrization of the CKM matrix. This matrix, in turn, can be transformed, up to phases, into another CKM matrix now parametrized by two $R_{12}$ matrices, two new complex conjugate phase matrices and a new $R_{23}(\beta_2)$ matrix. For the hierarchical case no new phases enter at this stage, while the angles $\beta_1$ and $\beta_3$ characterizing the $R_{12}$ matrices on the left and right, respectively, and the angle  $\beta_2$ in the $R_{23}$ matrix are given by:
\begin{equation}
\beta_1=\frac{\theta^d_{13}}{\phi_{23}}~~;~~\beta_2=\phi_{23}~~;~~\beta_{3}=-\frac{\theta^d_{13}}{\phi_{23}}.
\end{equation}
In arriving at these results we have assumed $\epsilon_u\sim\epsilon_d$ so that we could drop $\theta^u_{13}$ in comparison to $\theta^d_{13}$

\noindent {\it{ Step iii:}}

\noindent The two $R_{12}$ matrices on the left and right, along with some phases, can now be combined together into two other $R_{12}$ matrices. In the hierarchical limit, on the left, since $\beta_1\sim\epsilon$ while $ \theta^u_{12}\sim \epsilon^2$, the new angle $\rho_1$ is the same as the old angle $\beta_1$ ( $\rho_1=\beta_1$) and  no new phases enter. On the right-hand side, however, in this same limit a new phase $\eta_2$ and a new angle $\rho_3$ appear, with
\begin{equation}\rho_3e^{i\eta_2}=\theta^d_{12}+\beta_3e^{i(\delta_1-\gamma_2-\delta_3^d)}
\end{equation}
In addition, the resulting expression also contains a "CKM phase" $\lambda$ which, in the hierarchical limit, is given by:
\begin{equation}
\lambda=\gamma_2 -\delta_1+\delta^d_3 +\eta_2
\end{equation}

\noindent{\it{ Step iv:}}

\noindent In the final step the CKM matrix written in the parametrization with two $R_{12}$ matrices, a $R_{23}$ matrix,  and a CKM phase $\lambda$ is transformed into the desired "standard form" CKM matrix V. In this process, in the hierarchical case, a new phase $\kappa_4$ appears and the experimentally measured CKM phase $\delta$ is given by:
\begin{equation}
\delta=\lambda+\kappa_4=\gamma_2 -\delta_1+\delta^d_3 +\eta_2 +\kappa_4
\end{equation}
In the hierarchical case, two of the physically measured mixing angles, are given simply by:
\begin{equation}
\theta_{13}=\rho_1\beta_2=\beta_1\phi_{23}=\theta^d_{13}~~;~~\theta_{23}=\phi_{23}.
\end{equation}
The third mixing angle $\theta_{12}$ and the phase $\kappa_4$ are given by:
\begin{equation}
\theta_{12}e^{i\kappa_4}=\rho_3+\rho_1e^{-i\lambda}=\theta^d_{12}e^{-i\eta_2},
\end{equation}
where the 2nd equality follows from Eqs. (37) and (38). Thus $\kappa_4 +\eta_2=0$ and
\begin{equation}
\theta_{12}=\theta^d_{12}.
\end{equation}
From the above it follows that the CKM phase $\delta$ is given by:
\begin{equation}
\delta=\delta^d_3+\delta_{\rm{r}},
\end{equation}
where the residual phase $\delta_{\rm{r}}=\gamma_2-\delta_1$ can be inferred from the equation
\begin{equation}
\theta_{23}e^{i\delta_{\rm{r}}}=\theta^d_{23}-\theta^u_{23}e^{i(\delta_2-\delta_1)}.
\end{equation}

\noindent A few comments are in order:
\begin{enumerate}\item If $\theta^u_{23}$ could be neglected, the residual phase $\delta_{\rm{r}}$ vanishes and $\delta\to \delta^d_3$. This is easily understood, since effectively then $U^u\to P(0,\delta^u_1,\delta^u_2)$ and the CKM matrix is just $V(\theta_{23},\theta_{13},\theta_{12};\delta)\equiv  V(\theta^d_{23},\theta^d_{13},\theta^d_{12};\delta^d_3)$\\
\item Both the phase $\delta^d_3$ and the phase difference $\delta_2-\delta_1$ are directly related to the phases entering in the Yukawa matrices $Y^u$ and $Y^d$:
\begin{equation}
\delta^d_3=\gamma^d_{12} +\gamma^d_{23} -\gamma^d_{12}~~;~~\delta_2-\delta_1=\gamma^u_{23}-\gamma^d_{23}.
\end{equation}
\end{enumerate} 

\section{Discussion}

\noindent One sees from Eq. (46) that if the phases in the Yukawa matrices are of O(1), then so will be the phases $\delta^d_3$ and $\delta_{\rm{r}}$, and thus the CKM phase $\delta$ is also itself $\delta\sim O(1)$. This is totally consistent with the analysis of Heckman and Vafa.\cite{HV} However, our explicit calculation, keeping only the leading terms in $\epsilon$, suggests other possibilities. For instance, one could imagine that all the phases in the Yukawa matrices could be the same:
\begin{equation}
\gamma^u_{ij}=\gamma^d_{ij}= \delta_0 ~~(i\neq j)
\end{equation}
In this case, then $\delta_2-\delta_1=0$, so that $\delta=\delta^d_3$, and the experimentally measured CKM  phase is simply
\begin{equation}
\delta=\delta_0.
\end{equation}
Given our approximation of dropping subleading terms in $\epsilon$, a value of $\delta_0=\frac{\pi}{3}$ is perfectly compatible with the observed value of $\delta$ given in Eq. (10).

\noindent Obviously, the above discussion is very speculative and cannot be taken too seriously as an indication of some universal CP phase, which  imprints the Yukawa matrices. Indeed, the issue is much more complicated if one cannot argue, somehow, that the underlying theory yields Hermitian Yukawa matrices. As we mentioned earlier, by the polar decomposition theorem, \cite{FJ} a complex Yukawa matrix can be written as a product of a Hermitian matrix and a unitary matrix: $Y=Y_HU_R$. While one can always eliminate $U_R$ through a redefinition of the right-handed quarks, the phases in the Hermitian matrix $Y_H$ are not simply related to the phases in the general complex matrix Y. Thus, the idea of having a universal phase for the Hermitian Yukawa matrices may itself not be sensible. Nevertheless, we hope that the general considerations presented here may be useful in analyzing specific models for the Yukawa matrices.

\noindent As a final comment, it is worthwhile to note that in the process of passing from Yukawa matrices which are complex to Hermitian Yukawa matrices one performs a chiral transformation on the quarks by an angle Arg Det Y. This transformation changes the topological angle  $\theta$ which labels the QCD vacuum \cite{tH} into $\theta_{\rm{eff}}= \theta$ + Arg Det Y. \cite{JR} To avoid having an electric dipole moment for the neutron which is too large, the angle $\theta_{\rm{eff}} \leq 10^{-10}$. Why should this be so, is the strong CP problem. \cite{RDP} The F-theory example discussed suggests that naturally Arg Det Y $\sim O(1)$. Hence, to achieve $\theta_{\rm{eff}}<10^{-10}$ needs enormous fine tuning, unless some chiral symmetry, like that suggested long ago by Peccei and Quinn, \cite {PQ} efffectively drives $\theta_{\rm{eff}}\to 0$. 

\appendix
\section{  Diagonalization of Yukawa Matrices}
\noindent To diagonalize the Hermitian mass matrices $Y^u$ and $Y^d$ we will use an approach due to Rasin. \cite{Rasin} As a first step, it is useful  to transform them via a unitary transformation into real matrices with zeros in the 23 and 32 entries. Let us write:
\begin{equation}
Y=\left(\begin{array}{ccc}y_{11}&~y_{12}e^{i\gamma_{12}}&~~ y_{13}e^{i\gamma_{13}}\\y_{12}e^{-i\gamma_{12}}&y_{22}&~~y_{23}e^{i\gamma_{23}}\\y_{13}e^{-i\gamma_{13}}&~~y_{23}e^{-i\gamma_{23}}&y_{33} \end{array}\right)
\end{equation}
and
\begin{equation}
Y_{SF}= \left(\begin{array}{ccc}x_{11}&x_{12}&x_{13}\\x_{12}&x_{22}&0\\x_{13}&0&x_{33} \end{array}\right) .
\end{equation}
Then Y and $Y_{SF}$ are related by the unitary transformation
\begin{equation}
Y= U_{SF} Y_{SF}U_{SF}^{\dagger}
\end{equation}
with
\begin{equation}
U_{SF}= P(0,\gamma_{23},0)R_{23}(\theta'_{23})P(0,-\alpha_{12},-\alpha_{13})
\end{equation}
 where  
 \begin{equation}
 P(\alpha_1,\alpha_2,\alpha_3)= \left(\begin{array}{ccc}e^{i\alpha_1}&0&0\\0&e^{i\alpha_2}&0\\0&0&e^{i\alpha_3} \end{array}\right)
 \end{equation}
 and
 \begin{equation}
 R_{23}(\theta)=\left(\begin{array}{ccc}1&0&0\\0&cos \theta &\sin \theta\\0&-sin\theta &cos\theta. \end{array} \right)
 \end{equation}
 The relation between the parameters in Y and $Y_{SF}$ are given below:
 \begin{equation}
 \left(\begin{array}{c}x_{12}e^{i\alpha_{12}}\\x_{13}e^{i\alpha_{13}}\end{array}\right)=\left(\begin{array}{cc}\cos \theta_{23}'&\sin\theta_{23}' \\-sin \theta_{23}' & \cos \theta_{23}' \end{array}\right)\left(\begin{array}{c}y_{12}e^{i(\gamma_{12} +\gamma_{23})}\\y_{13}e^{i\gamma_{13}} \end{array}\right)
 \end{equation}
 and
 \begin{eqnarray}
 x_{11}&=&y_{11}\\
 x_{22}&=&\frac{1}{2}(y_{22} +y_{33}) -\frac{1}{2}\sqrt{(y_{33}-y_{22})^2 +4y_{23}^2}\\
 x_{33}&=&\frac{1}{2}(y_{22} +y_{33}) +\frac{1}{2}\sqrt{(y_{33}-y_{22})^2 +4y_{23}^2}\\
 \tan \theta_{23}'&=&\frac{y_{23}}{y_{33} -x_{22}}
 \end{eqnarray}
 
 \noindent The matrix $Y_{SF}$ can, in turn, be diagonalized by an orthogonal transformation
 \begin{equation}
 Y_{SF}= O_{SF}Y_{\rm{Diag}}O_{SF}^T
 \end{equation}
 Here
 \begin{equation}
 Y_{\rm{Diag}}=\left(\begin{array}{ccc}y_1&0&0\\0&y_2&0\\0&0&y_3\end{array}\right)
 \end{equation}
 and
 \begin{equation}
 O_{SF}=R_{23}(\theta_{23}'')R_{13}(\theta_{13})R_{12}(\theta_{12})
 \end{equation}
 where
 \begin{equation}
 R_{13}(\theta)=\left(\begin{array}{ccc}\cos \theta&0&\sin \theta\\0&1&0\\-\sin\theta&0 &\cos\theta \end{array} \right)~~;~~R_{12}(\theta)=\left(\begin{array}{ccc}\cos \theta&\sin \theta&0\\-\sin\theta&\cos\theta&0\\0&0&1 \end{array} \right).
 \end{equation}
 A straightforward calculation, using \cite{Rasin} $Y_{SF}O_{SF}=O_{SF}Y_{\rm{Diag}}$, yields for the mixing angles the formulas:
 \begin{equation}
 \tan\theta_{23}''=\frac{x_{12}}{x_{13}}\left[\frac{y_3-x_{33}}{y_3-x_{22}}\right]
 \end{equation}
 \begin{equation}
 \tan\theta_{13}= \frac{x_{12} \sin \theta_{23}''+x_{13}\cos\theta_{23}''}{y_3-x_{11}}
 \end{equation}
 \begin{equation}
 \tan \theta_{12}= \frac{\cos\theta_{13}[(y_1-x_{11}) +(y_3-x_{11})\tan^2\theta_{13}]}{x_{13} \sin\theta_{23}''-x_{12} \cos\theta_{23}''}
 \end{equation}
 
 \noindent It follows from the above that the unitary matrix $U=U_{SF}O_{SF}$ which diagonalizes Y is then
 \begin{equation}
 U= P(0,\gamma_{23},0)R_{23}(\theta'_{23})P(0,-\alpha_{12},-\alpha_{13})R_{23}(\theta_{23}'')R_{13}(\theta_{13})R_{12}(\theta_{12})
 \end{equation}
 The two $R_{23}$ rotations and the phase matrices can be combined together and one finds
 \begin{equation}
 P(0,\gamma_{23},0)R_{23}(\theta'_{23})P(0,-\alpha_{12},-\alpha_{13})R_{23}(\theta_{23}'')=P(0,\delta_1,\delta_2)R_{23}(\theta_{23})P(0,0,\delta_3)
 \end{equation}
 The angle $\theta_{23}$ and  two phases $\beta_{12}$ and $\beta_{13}$ can be determined from the equations
 \begin{eqnarray}
 \cos\theta_{23}e^{i\beta_{12}}&=& \cos \theta_{23}'\cos \theta_{23}'' e^{-i\alpha_{12}} -\sin \theta_{23}'\sin \theta_{23}'' e^{-i\alpha_{13}}\\\sin\theta_{23}e^{i\beta_{13}}&=& \cos \theta_{23}'\sin \theta_{23}'' e^{-i\alpha_{12}} +\sin \theta_{23}'\cos \theta_{23}'' e^{-i\alpha_{13}}.
 \end{eqnarray}
Then a simple calculation gives the following expression for the phases $\delta_i$ (i=1,2,3):
\begin{eqnarray}
\delta_1&=& \gamma_{23} +\beta_{12}\\
\delta_2&=& -\alpha_{12} -\alpha_{13}- \beta_{13}\\
\delta_3&=& \beta_{13}-\beta_{12}
\end{eqnarray} 
Since $P(0,0,\delta_3) R_{13}(\theta_{13})=P(0,0,\delta_3) R_{13}(\theta_{13})P(0,0,-\delta_3)P(0,0,\delta_3)$ while  $P(0,0,\delta_3) R_{12}(\theta_{12})=  R_{12}(\theta_{12})P(0,0, \delta_3)$ the matrix U can be written as
\begin{equation}
U=P(0,\delta_1,\delta_2)R_{23}(\theta_{23})P(0,0,\delta_3) R_{13}(\theta_{13})P(0,0,-\delta_3) R_{12}(\theta_{12})P(0,0, \delta_3)
\end{equation}
However, the last phase matrix on the right $P(0,0,\delta_3)$ can be dropped since it acts on $Y_{\rm{Diag}}$. Thus, effectively,
\begin{equation}
U=P(0,\delta_1,\delta_2)V(\theta_{23},\theta_{13},\theta_{12};\delta_3)
\end{equation}
where
\begin{equation}
V(\theta_{23},\theta_{13},\theta_{12};\delta_3)=R_{23}(\theta_{23})P(0,0,\delta_3) R_{13}(\theta_{13})P(0,0,-\delta_3) R_{12}(\theta_{12})
\end{equation}
is the CKM matrix \cite{CKM} written in the standard Chau-Keung parametrization. \cite{CK}
 
 \noindent In the hierarchical model considered in the text one can show that the quantity $y_3-x_{33}$ is very small [$y^u_3-x^u_{33}\sim O(\epsilon^8)$~;~~$y^d_3-x^d_{33}\sim O(\epsilon^5)$]. Thus, the angle $\theta_{23}''$ can be neglected altogether. Taking $\theta_{23}''=0$ and the leading order expressions for both $y_3$ and $y_2$ [$y_3=x_{33}=y_{33}=1$ and $y_2=x_{22}=y_{22}-y_{23}^2$] it follows that
 \begin{equation}
 \tan \theta_{13}= x_{13}
 \end{equation}
 Since $\rm{Det}Y_{SF}=\rm{Det} Y_{\rm{Diag}}$,  in leading order one finds
 \begin{equation}
 y_1=x_{11}- x_{13}^2 -\frac{x^2_{12}}{x_{22}}.
 \end{equation}
 Thus
 \begin{equation}
 \tan\theta_{12}= -\frac{y_1-x_{11} +\tan^2\theta_{13}}{x_{12}}= \frac{x_{12}}{x_{22}}
 \end{equation}
 In the limit that $\theta_{23}''\to 0$ clearly $\theta_{23}=\theta_{23}'$, so that in leading order 
 \begin{equation}
 \tan\theta_{23}=y_{23}.
 \end{equation}
 In this approximation, the phases $\beta_{12}$ and $\beta_{13}$ are simply $ \beta_{12}=-\alpha_{12}$ and $\beta_{13}=-\alpha_{13}$. Hence the phases entering in U are:
 \begin{eqnarray}
  \delta_1&=& \gamma_{23} -\alpha_{12}\\
\delta_2&=& -\alpha_{12} \\
\delta_3&=& \alpha_{12}-\alpha_{13}
\end{eqnarray}  

\section{ Computation of the CKM Matrix}
\noindent The Hermitian Yukawa matrices for the u- and d-quarks, $Y^u$ and $Y^d$, are diagonalized by the unitary matrices $U^u$ and $U^d$, respectively:
\begin{equation}
U^{u\dagger}Y^uU^u=Y^u_{\rm{Diag}}~~;~~U^{d\dagger}Y^dU^d=Y^d_{\rm{Diag}}
\end{equation}
As shown in Appendix A, the unitary matrices $U^i$ $(i=u,d)$ have the form
\begin{equation}
U^i= P(0,\delta^i_1, \delta^i_2) V(\theta^i_{23},\theta^i_{13},\theta^i_{12};\delta^i_3),
\end{equation}
where V is the CKM matrix written in "standard form":
\begin{equation}
V(\theta^i_{23},\theta^i_{13},\theta^i_{12};\delta^i_3)= R_{23}(\theta^i_{23})P(0,0,\delta^i_3)R_{13}(\theta^i_{13})P(0,0,-\delta^i_3)R_{12}(\theta^i_{12}).
\end{equation}
The CKM matrix itself can be computed from the product of $U^{u\dagger}$ and $U^d$ and can be put in "standard form" after removing five unphysical phases. That is,
\begin{equation}
U^{u\dagger}U^d= P(0,\alpha_1,\alpha_2)U_{\rm{CKM}}(\theta_{23},\theta_{13},\theta_{12};\delta)P(\alpha_3,\alpha_4,\alpha_5)
\end{equation}
with
\begin{equation}
U_{\rm{CKM}}(\theta_{23},\theta_{13},\theta_{12};\delta)=V(\theta_{23},\theta_{13},\theta_{12};\delta)
\end{equation}

\noindent
To arrive at the above result requires a series of manipulations by means of which  one computes the three angles $\theta_{23},\theta_{13},\theta_{12}$ and the phase $\delta$ in terms of the six angles $\theta^i_{23},\theta^i_{13},\theta^i_{12}$ ($i=u,d$) and the four phases $\delta_1=\delta_1^d-\delta_1^u,\delta_2=\delta_2^d-\delta_2^u,\delta_3^d,$ and $\delta_3^u$  in $U^u$ and $U^d$.

\noindent In detail one has
\begin{eqnarray}
U^{u\dagger}U^d&=& V^{\dagger}(\theta^u_{23},\theta^u_{13},\theta^u_{12};\delta^u_3)P(0,-\delta^u_1, -\delta^u_2)P(0,\delta^d_1, \delta^d_2)V(\theta^d_{23},\theta^d_{13},\theta^d_{12};\delta^d_3)\nonumber \\~~~~~&=&R_{12}^T(\theta^u_{12})P(0,0,\delta^u_3)R_{13}^T(\theta^u_{13})P(0,0,-\delta^u_3)R_{23}^T(\theta^u_{23})P(0,\delta_1,\delta_2) \nonumber \\~~~~~~&\times &R_{23}(\theta^d_{23})P(0,0,\delta^d_3)R_{13}(\theta^d_{13})P(0,0,-\delta^d_3)R_{12}(\theta^d_{12})
\end{eqnarray}
To transform this expression into the desired form we go through a series of steps, employing a number of identities for CKM matrices derived by Rasin. \cite{Rasin} As a first step, we combine the two  $R_{23}$ matrices and the central phase matrix above into another $R_{23}$ matrix and two phase matrices:
\begin{equation}
M_1=R_{23}^T(\theta^u_{23})P(0,\delta_1,\delta_2)R_{23}(\theta^d_{23})=P(0,0,\delta_1+\delta_2-\gamma_1-\gamma_2)R_{23}(\phi_{23})P(0,\gamma_1,\gamma_2)
\end{equation}
where the angle $\phi_{23}$ and phases $\gamma_1$ and $\gamma_2$ can be computed from the equations
\begin{eqnarray}
\cos\phi_{23}e^{i\gamma_1}&=& \cos\theta^u_{23}\cos\theta^d_{23}e^{i\delta_1} +\sin\theta^u_{23}\sin\theta^d_{23}e^{i\delta_2} \nonumber\\
\sin\phi_{23}e^{i\gamma_2}&=& \cos\theta^u_{23}\sin\theta^d_{23}e^{i\delta_1} -\sin\theta^u_{23}\cos\theta^d_{23}e^{i\delta_2}
\end{eqnarray}

\noindent As a second step we combine the above result with the two phase matrices and the two $R_{13}$ matrices. It is useful to define
\begin{equation}
\omega=\delta_1+\delta_2-\gamma_1+\delta^d_3 -\delta^u_3~~;~~ \tau=\gamma_2+\delta^d_3
\end{equation}
Then, after some rearrangement of the phase matrices, one can write
\begin{eqnarray}
M_2&=&R_{13}^T(\theta^u_{13}) P(0,0,-\delta^u_3)M_1P(0,0,\delta^d_3)R_{13}(\theta^d_{13}) \nonumber \\
&=& P(0,\tau,\omega)M_{13}P(0,\gamma_1-\tau,0)
\end{eqnarray}
where
\begin{equation}
M_{13}=P(\omega,0,0)R_{13}^T(\theta^u_{13})P(-\omega,0,0)R_{23}(\phi_{23})R_{13}(\theta^d_{13})
\end{equation}
The above matrix corresponds to a particular parametrization of the CKM matrix. Using one of the identities of Rasin \cite{Rasin} it can be related to another CKM matrix in a parametrization involving two $R_{12}$ matrices and an $R_{23}$ matrix. One introduces in this way three new angles $\beta_1,\beta_2$ and $\beta_3$ and three new phases $\xi_1,\xi_2$ and $\sigma$ related to the angles and phase in $M_{13}$ In detail, one has:
\begin{equation}
 M_{13}=P(0,\xi_1,\xi_2)M_{12}P(0,-\xi_2,-\xi_1)
 \end{equation}
 where
 \begin{equation}
 M_{12}= P(-\sigma,0,0)R_{12}(\beta_1)P(\sigma,0,0)R_{23}(\beta_2)R_{12}(\beta_3),
 \end{equation}
  and one identifies
 \begin{eqnarray}
 \cos\beta_2e^{i(\xi_2-\xi_1)}&=& \sin\theta^u_{13}\sin\theta^d_{13}e^{-i\omega} +\cos\theta^u_{13}\cos\theta^d_{13}\cos\phi_{23} \nonumber \\
 \cos\beta_1\sin\beta_2&=& \sin\phi_{23}\cos\theta^d_{13} \nonumber\\
 \sin\beta_1\sin\beta_2e^{-i(\sigma+\xi_1)}&=&\cos\theta^u_{13}\sin\theta^d_{13} -\sin\theta^u_{13}\cos\theta^d_{13}\cos\phi_{23}e^{i\omega} \nonumber \\
 \cos\beta_3\sin\beta_2&=& \sin\phi_{23}\cos\theta^u_{13} \nonumber\\
 \sin\beta_3\sin\beta_2e^{i \xi_2}&=&\sin\theta^u_{13}\cos\theta^d_{13}e^{-i\omega} -\cos\theta^u_{13}\sin\theta^d_{13}\cos\phi_{23}
 \end{eqnarray}
 
\noindent As a third step, in the expression for $U^{u\dagger}U^d$ one combines the $R_{12}$ matrices on the right and left into new $R_{12}$ matrices. Dropping irrelevant phase matrices on the far left and  far right in  $U^{u\dagger}U^d$ and combining the other phase matrices appropriately, one arrives at the following expression for $U^{u\dagger}U^d$:
 \begin{eqnarray}
 U^{u\dagger}U^d &=&[R^T_{12}(\theta^u_{12}) P(-\sigma,\tau+\xi_1,0)R_{12}(\beta_1)]P(\sigma,0,0)R_{23}(\beta_2) \nonumber \\& \times &[R_{12}(\beta_3) P(0,\gamma_1-\tau-\xi_2,0)R_{12}(\theta^d_{12})]\nonumber\\&=&M_LP(\sigma,0,0)R_{23}(\beta_2)M_R
 \end{eqnarray}
 It is straightforward to work out the structure of $M_L$ and $M_R$. One finds
 \begin{equation}
 M_L=P(\epsilon_2,\tau+\xi_1-\sigma-\epsilon_1,0)R_{12}(\rho_1)P(\epsilon_1-\epsilon_2,0,0)
 \end{equation}
 where
 \begin{eqnarray}
 \cos\rho_1e^{i\epsilon_1}&=& \cos\theta^u_{12}\cos\beta_1e^{-i\sigma} +\sin\theta^u_{12}\sin\beta_1e^{i(\tau +\xi_1)} \nonumber \\
\sin\rho_1e^{i\epsilon_2}&=& \cos\theta^u_{12}\sin\beta_1e^{-i\sigma} -\sin\theta^u_{12}\cos\beta_1e^{i(\tau +\xi_1)}.
\end{eqnarray}
For $M_R$ the result is:
\begin{equation}
M_R=P(\eta_1+\eta_2+\tau+\xi_2-\gamma_1,0,0)R_{12}(\rho_3)P(-\eta_2+\gamma_1-\tau-\xi_2,-\eta_1+\gamma_1-\tau-\xi_2,0)
\end{equation}
where
\begin{eqnarray}
\cos\rho_3e^{i\eta_1}&=&\cos\beta_3\cos\theta^d_{12}- \sin\beta_3\sin\theta^d_{12}e^{i(\gamma_1-\tau-\xi_2)} \nonumber\\
\sin\rho_3e^{i\eta_2}&=&\cos\beta_3\sin\theta^d_{12}+ \sin\beta_3\cos\theta^d_{12}e^{i(\gamma_1-\tau-\xi_2)}
\end{eqnarray}
Dropping the irrelevant phase matrices on the left and right, and combining the phase matrices in the middle, the above results yield an expression for $U^{u\dagger}U^d$ which is of the CKM form:
\begin{equation}
U^{u\dagger}U^d=P(-\lambda,0,0)R_{12}(\rho_1)P(\lambda,0,0)R_{23}(\beta_2)R_{12}(\rho_3)
\end{equation}
where the phase $\lambda$ is given by:
\begin{equation}
\lambda=\sigma+\epsilon_1-\epsilon_2 +\eta_1+\eta_2+\tau+\xi_2-\gamma_1
\end{equation}

\noindent As a fourth and final step, one needs to transform the expression for $U^{u\dagger}U^d$ above into the CKM "standard form" V. For this purpose one can use another Rasin identity \cite{Rasin} to relate the two CKM matrices. Starting from the identity
\begin{equation}
P(-\lambda,0,0)R_{12}(\rho_1)P(\lambda,0,0)R_{23}(\beta_2)R_{12}(\rho_3)=P(0,\kappa_1,\kappa_2)V(\theta_{23},\theta_{13},\theta_{12};\delta)P(\kappa_3,\kappa_4,\kappa_5)
\end{equation}
a straightforward calculation identifies three of the five phases $\kappa_i$ and the CKM phase $\delta$ as:
\begin{equation}
-\kappa_1=-\kappa_2=\kappa_5=\kappa_3 +\kappa_4~~;~~ \delta=\lambda+\kappa_3+\kappa_4
\end{equation}
The remaining two phases $\kappa_3$ and $\kappa_4$, as well as the three CKM angles $\theta_{23}, \theta_{13}$ and $\theta_{12}$, can be obtained from the following equations:
\begin{eqnarray}
\sin\theta_{13}&=&\sin\rho_1\sin\beta_2 \nonumber \\
\sin\theta_{23}\cos\theta_{13}&=&\cos\rho_1\sin\beta_2 \nonumber \\
\cos\theta_{23}\cos\theta_{13}&=&\cos\beta_2 \nonumber \\
\cos\theta_{12}\cos\theta_{13}e^{i\kappa_3}&=&\cos\rho_1\cos\rho_3 -\sin\rho_1\sin\rho_3\cos\beta_2e^{-i\lambda} \nonumber \\
\sin\theta_{12}\cos\theta_{13}e^{i\kappa_4}&=&\cos\rho_1\sin\rho_3 +\sin\rho_1\cos\rho_3\cos\beta_2e^{-i\lambda}
\end{eqnarray}

\noindent The above results are general and entail no approximations. The results, however, simplify considerably when  the Yukawa matrices have a hierarchical structure. In the particular case considered in the text, the angles in the unitary matrices which diagonalize the Yukawa matrices have the following hierarchical pattern:
\begin{equation}
\theta^u_{12}=O(\epsilon^2), \theta^u_{23}=O(\epsilon^2), \theta^u_{13}=O(\epsilon^4); \theta^d_{12}=O(\epsilon), \theta^d_{23}=O(\epsilon^2), \theta^d_{13}=O(\epsilon^3).
\end{equation}
Given these hierarchies, as outlined in the text, a simple calculation shows that the angles $\theta_{12}$ and $\theta_{13}$ are given by:
\begin{equation}
\theta_{12}=\theta^d_{12}~~;~~\theta_{13}=\theta^d_{13},
\end{equation}
while $\theta_{23}$ and the residual phase $\delta_{\rm{r}}=\gamma_2-\delta_1$  obey
\begin{equation}
\theta_{23}e^{i\delta_{\rm{r}}}=\theta^d_{23}-\theta^u_{23}e^{i(\delta_2-\delta_1)}.
\end{equation}
The CKM phase is given by
\begin{equation}
\delta=\delta^d_3+\delta_{\rm{r}}
\end{equation}
As we discussed in Appendix A, in the hierarchical case, the angles $\theta^u_{23}$ and $\theta^d_{23}$ are simply
\begin{equation}
\theta^u_{23}=y^u_{23}~~;~~ \theta^d_{23}=y^d_{23},
\end{equation}
while the phases $\delta^d_3$ and $\delta_2-\delta_1$ are given by:
\begin{equation}
\delta^d_3=\alpha^d_{12}-\alpha^d_{13}=\gamma^d_{12}+\gamma^d_{23}-\gamma^d_{13}~~;~~
\delta_2-\delta_1= \delta^d_2-\delta_1^d-\delta^u_2+\delta^u_1=\gamma^u_{23}-\gamma^d_{23}
\end{equation}

\end{document}